# Thermal Coherence of Heisenberg Model With Dipole Interaction and Magnetic External Field

Jin-Kai Li,  Guo-Feng Zhang[*]

*School of Physics, Beihang University, Beijing 100191, China*

**Abstract**：The dynamics of coherence in a two-qubit Heisenberg XXX model with magnetic field in the presence of the intrinsic decoherence is investigated. We discuss the influence of dipole parameter $D$, spin distance $r$, magnetic field $B_z$ and initial state parameter $\alpha$ on coherence of the long time limit situation. Then, we discuss how decoherence changes with time with the initial state parameter $\alpha$ and the decoherence parameter $\gamma$.

## I.  INTRODUCTION

Coherence is a very important concept in quantum mechanics. Coherence is initially introduced from interference phenomena in wave optics, such as double-slit interference experiment, which is impossible in ray optics. Since quantum mechanics is risen as a unified picture of waves and particles coherence has been widely discussed. Coherence has a close relationship with entanglement which plays a central role in applications of quantum physics and quantum information science. As follows with the assistant of measurement disposal, we discuss the violation between our everyday classical perceptions and the predictions of the Schrödinger equation which a state evolves into a superposition state that simultaneously contains many alternatives to better understand coherence.

The explanations of measurement such as the Copenhagen interpretation proposed by Niels Bohr and the many-worlds interpretation developed in the 1950s by Hugh Everett Ⅲ, we do not discuss them in detail here [1][6]. A convenient starting point for the discussion of the measurement problem, and more generally of the emergence of classical behavior from quantum dynamics, is the analysis of quantum measurements due to John von Neumann [7]. Briefly speaking, he uses reduced density matrix which cancels the off-diagonal terms in density matrix other than density matrix to describe a completed quantum measurement. Because the advantage of reduced density matrix over density matrix is that its coefficients may be interpreted as classical probabilities. For example, we detect a two-state system and the detector is also two-state. In density matrix case, the state can be understood as spin along z direction also and x direction, but when the off-diagonal terms are absent, one can safely maintain the system and apparatus are each separately in a definite but unknown state [8][10]. Reduced density matrix can be also obtained from tracing over "the environment" degrees of freedom to maintain the system-detector combination as in information processing. Now, a superposition of the records-of the states of detector – is no longer a record of a superposition. Any coherent superposition states of the detector is continuously reduced to a mixture.

---

[*] Corresponding author. gf1978zhang@buaa.edu.cn

In the real world, the spreading of quantum correlations is practically inevitable which will result in a correlation with the environment and will necessarily lead to decoherence. In mathematics, it means the density matrix of the system and detector pair will lose its off-diagonal system in contact with environment, which is known as the process by which quantum information is discarded. In the understanding of quantum fundamentals, particularly quantum measurement and the emergence of classicality in quantum systems, decoherence always plays a crucial role [11]. In addition, the understanding of decoherence is essential in technologies that actively use quantum coherence. Milburn changed the Schrödinger equation slightly in order to show coherence is damaged as the system evolves in Ref. [12]. This type of decoherence is often called intrinsic decoherence. Recently, the impression of intrinsic decoherence has been used in open system [13], quantum gate [14] and various Heisenberg chains [15][20].

In this paper, decoherence of Heisenberg XXX model with magnetic field is discussed in detail. Analytically deriving coherence of two-qubit state allows to focus on the effect of dipolar interaction ($D$). Findings reveal that $D$ and $B_z$ are decisive parameters on decoherence. Spin distance $r$ and initial state parameter $\alpha$ on coherence of the long time limit situation are also discussed. Then, we discuss how decoherence changes with time with the initial state parameter $\alpha$ and the intrinsic decoherence parameter $\gamma$.

## II. Description of the Model and the Definition of Decoherence

Heisenberg spin chain is a model used for discussing the ferromagnetic property of matter. Ferromagnetism is originated in parallel spin array of material. Heisenberg raised this model based in the quantum mechanics which proposes ferromagnetic spontaneous magnetization is originated in electric exchange interaction between electrics. Heisenberg exchange interaction model proposed that there is exchange interaction between atoms in magnetic substance, and this interaction is merely between the nearest atoms. The Hamiltonian of the N-qubit one-dimensional spin chain with dipole-dipole interaction term and magnetic field along z direction is risen as,

$$\hat{H} = -\frac{1}{2}\sum_{i=1}^{N}[J_x\hat{\sigma}_i^x\hat{\sigma}_{i+1}^x + J_y\hat{\sigma}_i^y\hat{\sigma}_{i+1}^y + J_z\hat{\sigma}_i^z\hat{\sigma}_{i+1}^z + \frac{B_z}{2}(\hat{\sigma}_i^z + \hat{\sigma}_{i+1}^z)] + \hat{H}_D ,  \quad (1)$$

$$\hat{H}_D = \frac{D}{2}\sum_{i=1}^{N}[\frac{\hat{\sigma}_i\hat{\sigma}_{i+1}}{r_{i,i+1}^3} - 3\frac{(\hat{\sigma}_i \cdot r_{i,i+1})(\hat{\sigma}_{i+1} \cdot r_{i,i+1})}{r_{i,i+1}^5}] , \quad (2)$$

where $J_x$, $J_y$ and $J_z$ are exchange coupling constants and negative (positive) values are corresponds to FM (AFM) interaction, $B_z$ denotes magnetic field along z-axis and $D$ represents the strength of dipole interaction ($r_{i,j}$:two-spin distance vector). Considering only nearest neighboring spin interactions, $\hat{H}_D$ changed into a more simple type as expressed in Eq. (2) in which periodic boundary conditions (PBC) are satisfied $\hat{\sigma}_1 = \hat{\sigma}_{N+1}$. A clear illustration of the model is given in Fig. 1.

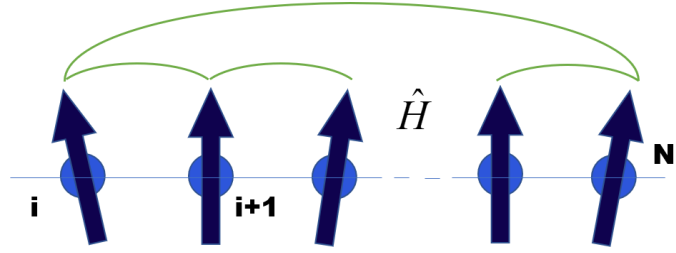

Fig. 1: Illustration of spin chain: Green narrow and wide rounded lines represent the interacted spin pairs and PBC respectively. In this paper, we only consider the situation for N=2.

In case of spin chain including only nearest-neighbor dipolar interaction term, we can simplify Hamiltonian of system as expressed in Eq. (3) where $r$ denotes distance between spins.

$$\hat{H} = -(\frac{J_x}{2} + \frac{D}{r^3})\hat{S}_1^x\hat{S}_2^x + (\frac{D}{2r^3} - \frac{J_y}{2})\hat{S}_1^y\hat{S}_2^y + (\frac{D}{2r^3} - \frac{J_z}{2})\hat{S}_1^z\hat{S}_2^z - \frac{B_z}{2}(\hat{S}_1^z + \hat{S}_2^z) \ . \quad (3)$$

For further simplicity, we use methods as follows. $\hat{\sigma}_m^n$ (where $m$:1,2 and $n$:+,-) are known as Pauli spin matrices operating to up/down spin in the basis $|\uparrow\rangle$, $|\downarrow\rangle$, where $|\uparrow\rangle = (1 \ 0)^T$, $|\downarrow\rangle = (0 \ 1)^T$, $T$ represents matrix transposition. Raising and lowering operators are $\hat{\sigma}^\pm = \frac{1}{2}(\hat{\sigma}^x \pm i\hat{\sigma}^y)$, then the Hamiltonian is simplified as

$$\hat{H} = (-\frac{3D}{2r^3} - \frac{J_x}{2} + \frac{J_y}{2})(\hat{\sigma}_1^+\hat{\sigma}_2^+ + \hat{\sigma}_1^-\hat{\sigma}_2^-) + (-\frac{D}{2r^3} - \frac{J_x}{2} - \frac{J_y}{2})(\hat{\sigma}_1^+\hat{\sigma}_2^- + \hat{\sigma}_1^-\hat{\sigma}_2^+) \\ + (\frac{D}{2r^3} - \frac{J_z}{2})\hat{\sigma}_1^z\hat{\sigma}_2^z - \frac{B_z}{2}(\hat{\sigma}_1^z + \hat{\sigma}_2^z) \quad . \quad (4)$$

Table 1: Eigenvalues and corresponding eigenvectors of $\hat{H}$.

| Eigenvalues | Eigenvectors |
| --- | --- |
| $E_1 = \dfrac{3J}{2}$ | $\|\phi_1\rangle = \dfrac{1}{\sqrt{2}}(-|\uparrow\downarrow\rangle + |\downarrow\uparrow\rangle)$ |
| $E_2 = \dfrac{-2D - Jr^3}{2r^3}$ | $\|\phi_2\rangle = \dfrac{1}{\sqrt{2}}(|\uparrow\downarrow\rangle + |\downarrow\uparrow\rangle)$ |
| $E_3 = \dfrac{D - J \cdot r^3 - \sqrt{9D^2 + 4B_z^2 r^6}}{2r^3}$ | $\|\phi_3\rangle = \eta^+(\delta|\uparrow\uparrow\rangle + |\downarrow\downarrow\rangle)$ |
| $E_4 = \dfrac{D - J \cdot r^3 + \sqrt{9D^2 + 4B_z^2 r^6}}{2r^3}$ | $\|\phi_4\rangle = \eta^-(\varepsilon|\uparrow\uparrow\rangle + |\downarrow\downarrow\rangle)$ |

Matrix form of the $\hat{H}$ is given in Eq. (5) in which we only deal with the isotropic ferromagnetic state where $J_\alpha = J$.

$$\hat{H} = \begin{pmatrix} -B_z - \frac{J}{2} + \frac{D}{2r^3} & 0 & 0 & -\frac{3D}{2r^3} \\ 0 & \frac{J}{2} - \frac{D}{2r^3} & -J - \frac{D}{2r^3} & 0 \\ 0 & -J - \frac{D}{2r^3} & \frac{J}{2} - \frac{D}{2r^3} & 0 \\ -\frac{3D}{2r^3} & 0 & 0 & B_z - \frac{J}{2} + \frac{D}{2r^3} \end{pmatrix}. \tag{5}$$

Eigenvalues and corresponding eigenvectors of the Hamiltonian (Eq.(5)) are given in Table 1 where $\eta^+ = (\delta^2 + 1)^{-\frac{1}{2}}$ and $\eta^- = (\varepsilon^2 + 1)^{-\frac{1}{2}}$ are normalization constants, while $\delta = \frac{2B_z \cdot r^3 + \tau}{3D}$ and $\varepsilon = \frac{2B_z \cdot r^3 - \tau}{3D}$, where $\tau = \sqrt{9D^2 + 4B_z^2 r^6}$.

The master equation in Milburn's model describing the intrinsic decoherence under the Markovian approximation, developed by the assumption that in adequately short time steps the system does not change continuously under unitary evolution but particularly in a random sequence of identical unitary transformations, becomes [21]

$$\frac{d\hat{\rho}(t)}{dt} = -i[\hat{H}, \hat{\rho}(t)] - \frac{\gamma}{2}[\hat{H},[\hat{H}, \hat{\rho}(t)]]. \tag{6}$$

The reduced Planck constant is set as unity, $\hbar = 1$, and the intrinsic decoherence parameter $\gamma$ is set as a constant minimum time step. In the limit $\gamma \to 0$, Eq.(6) reduces to the standard von Neumann equation while large $\gamma$ makes the evolution deteriorate from the unitary one. The above master equation can be solved as [22][23]

$$\hat{\rho}(t) = \sum_{k=0}^{\infty} \frac{(\gamma t)^k}{k!} \hat{M}^k(t) \hat{\rho}(0) \hat{M}^{\dagger k}(t), \tag{7}$$

where

$$\hat{M}^k(t) = \hat{H}^k \exp(-i\hat{H}t) \exp(-\gamma \hat{H}^2 t/2), \tag{8}$$

and $\hat{\rho}(0)$ is the initial state. By using the complete relation $\sum_n |\phi_n\rangle\langle\phi_n| = 1$ in above equation, the time evolution of the density matrix is given by

$$\hat{\rho}(t) = \sum_{m,n} \exp\left[-\frac{\gamma t}{2}(E_m - E_n)^2 - i(E_m - E_n)t\right] \times \langle\phi_m|\hat{\rho}(0)|\phi_n\rangle |\phi_m\rangle\langle\phi_n|. \tag{9}$$

In Eq.(9), $E_m$ and $|\phi_m\rangle$ are the eigenvalues and the corresponding eigenstates of $\hat{H}$ already given in Table 1.

After introducing the Heisenberg spin chain and the dynamical model, we turn our attention to the definition of decoherence below.

To better understand decoherence, we should consult quantum optical methods by which coherence is studied in terms of phase space distributions and multipoint correlation functions to provide a framework that concerns classical electromagnetic phenomena [24][25]. This is useful in getting intuition from classical wave mechanics and identifying those parts for which quantum coherence deviates from classical coherence. As a result, a wide variety of measures of coherence is in use (often functions of a density matrix' off-diagonal entries) whose use tends to be justified principally on the grounds of physical intuition. In our work, we investigate the quantum coherence of the system at finite (non-zero) temperature and establish a quantitative theory of coherence as the distance between the density matrix $\hat{\rho}$ and the decohered density matrix $\hat{\rho}_d$. The decohered density matrix is described according to the matrix elements

$$\langle \hat{\sigma}_1 \hat{\sigma}_2 | \hat{\rho}_d | \hat{\sigma}_1' \hat{\sigma}_2' \rangle = \langle \hat{\sigma}_1 \hat{\sigma}_2 | \hat{\rho} | \hat{\sigma}_1' \hat{\sigma}_2' \rangle \delta_{\hat{\sigma}_1, \hat{\sigma}_1'} \delta_{\hat{\sigma}_2, \hat{\sigma}_2'}. \tag{10}$$

Ref.[26] provides two theories of coherence namely the relative entropy measure belonging to the entropic class and the $\ell_1$-norm of coherence which is a geometric measure. We choose another measure based on the quantum version of the Jensen-Shannon divergence [27][27][29] explained in Ref. [30]

$$D(\hat{\rho}, \hat{\sigma}) \equiv \sqrt{J(\hat{\rho}, \hat{\sigma})} = \sqrt{S\left(\frac{\hat{\rho} + \hat{\sigma}}{2}\right) - \frac{S(\hat{\rho})}{2} - \frac{S(\hat{\sigma})}{2}}. \tag{11}$$

Here $\hat{\rho}$, $\hat{\sigma}$ are arbitrary density matrices and $S = -\text{Tr}\hat{\rho}\log\hat{\rho}$ is the von Neumann entropy. This measure has the advantage that it has both geometric and entropic properties and is also a metric satisfying triangle inequality. The coherence is then defined in our case as

$$C(\hat{\rho}) = D(\hat{\rho}, \hat{\rho}_d). \tag{12}$$

The coherence of the two-spin XXX model with dipole interaction in magnetic field in Eq. (5) is discussed in the rest of the paper.

### III. Decoherence

In order to get the state at any time $t$, we need the initial state. Without loss of generality, $\hat{\rho}(0) = \cos^2\alpha |\psi^+\rangle\langle\psi^+| + \sin^2\alpha |\phi^+\rangle\langle\phi^+|$ is chosen, where $\alpha \subset [0,\pi]$. $|\psi^+\rangle = (1/\sqrt{2})(|\downarrow\downarrow\rangle + |\uparrow\uparrow\rangle)$ and $|\phi^+\rangle = (1/\sqrt{2})(|\downarrow\uparrow\rangle + |\uparrow\downarrow\rangle)$ are the Bell states. The initial state is a mixed state.

Substituting $\rho(0)$ into Eq. (9), one can obtain the nonzero elements of the evolved density matrix

$$\rho_{11} = \frac{(m^2 - B_z \cdot \tau + B_z \cdot \tau \cdot e^{-2m^2 t\gamma} \cdot \cos(2mt))\cos^2\alpha}{2m^2},$$

$$\rho_{41} = \rho_{14}^* = \frac{e^{-2t(im+m^2\gamma)} \cdot (2\tau^2 e^{2t(im+m^2\gamma)} + B_z^2 + B_z \cdot (-e^{4imt} \cdot (-B_z + m) + m)) \cdot \cos^2\alpha}{4m^2}, \quad (13)$$

$$\rho_{44} = \frac{(m^2 + B_z \cdot \tau - B_z \cdot \tau \cdot e^{-2m^2 t\gamma} \cdot \cos(2mt))\cos^2\alpha}{2m^2},$$

$$\rho_{22} = \rho_{33} = \rho_{23} = \rho_{32} = \frac{\sin^2\alpha}{2}.$$

First of all, the long time limit situation is considered. As the time $t$ approaches infinity, the nonzero elements of the stationary state are

$$\rho_{11} = \frac{(m^2 - B_z \cdot \tau) \cdot \cos^2\alpha}{2m^2},$$

$$\rho_{41} = \rho_{14} = \frac{2\tau^2 \cdot \cos^2\alpha}{4m^2},$$

$$\rho_{44} = \frac{(m^2 + B_z \cdot \tau) \cdot \cos^2\alpha}{2m^2}, \quad (14)$$

$$\rho_{22} = \rho_{33} = \rho_{23} = \rho_{32} = \frac{\sin^2\alpha}{2}.$$

According to Eq. (11) and using Eq. (14) and corresponding $\hat{\rho}_d$, the long time limit coherence $C(\hat{\rho}) = D(\hat{\rho}, \hat{\rho}_d)$, also known as the steady state coherence, can be obtained. Afterwards we can get the characteristics of decoherence. However, as the expression is lengthily and not very informative, we do not write them explicitly here. By calculation we can plot the steady state coherence to make a mathematical analysis.

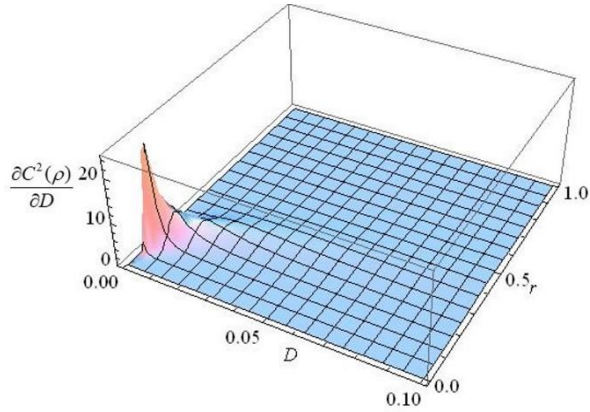

Fig. 2: D and r dependence of $\partial C^2(\rho)/\partial D$ with $\alpha = \pi/3$ and $B_z = 1$.

Because the expression of coherence $C(\rho)$ is after a radical sign, it is not convenience to discuss the change for $C(\rho)$ and the radical sign does not change monotonicity, we can investigate the coherence $C(\rho)$ squared, marked $C^2(\rho)$, to simplify the question. $\frac{\partial C^2(\rho)}{\partial D}$, the partial derivative of $C^2(\rho)$ to $D$ when initial parameter $\alpha = \pi/3$ and $B_z = 1$ is shown in Fig.2.

From Fig.2, there is always bigger than zero for the partial derivative, so $C^2(\rho)$, as well as $C(\rho)$, is monotone increasing with the dipole parameter $D$.

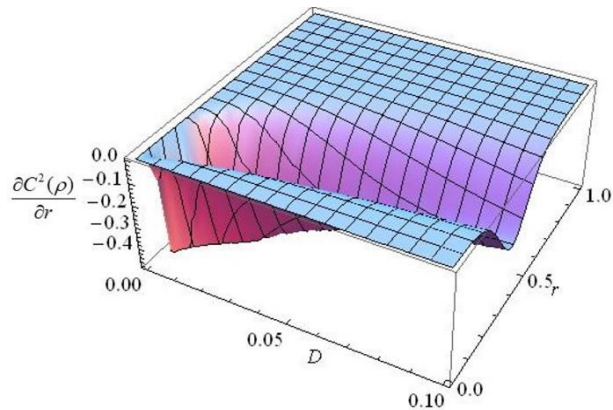

Fig. 3: D and r dependence of $\partial C^2(\rho)/\partial r$ with $\alpha = \pi/3$ and $B_z = 1$.

$\dfrac{\partial C^2(\rho)}{\partial r}$, the partial derivative of $C^2(\rho)$ to $r$ when initial parameter $\alpha = \pi/3$ and $B_z = 1$ is shown in Fig.3. From Fig.3, there is always smaller than zero for the partial derivative, so $C^2(\rho)$, as well as $C(\rho)$, is monotone decreasing with the distance parameter $r$.

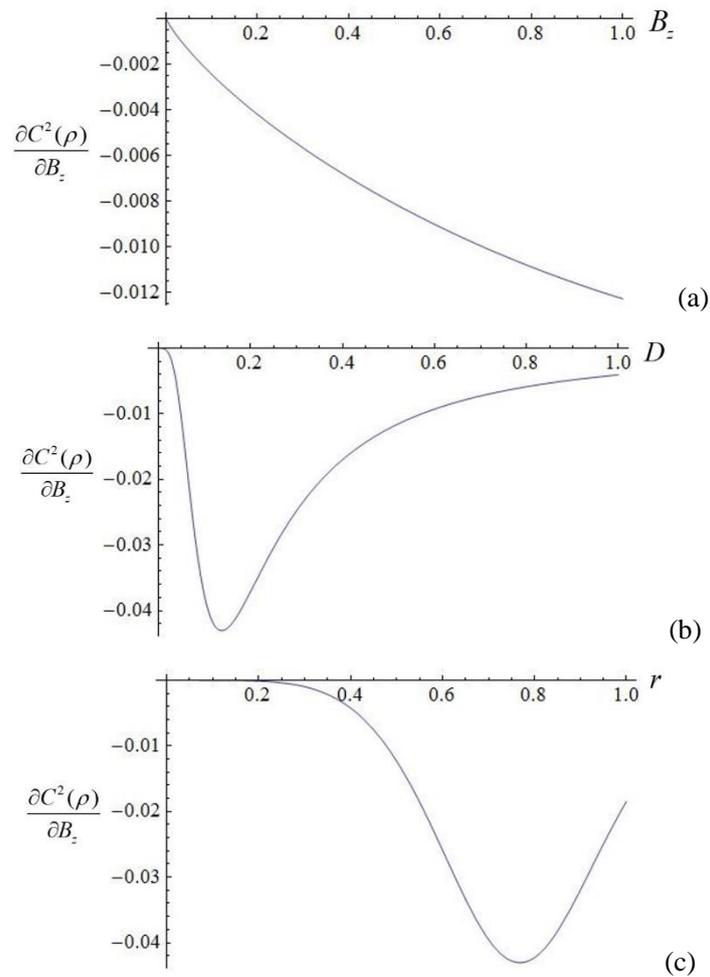

Fig. 4: (a) The change of $\partial C^2(\rho)/\partial B_z$ with $B_z$ when $\alpha = \pi/3$, $D = 0.5$ and $r = 0.5$. (b) The change of $\partial C^2(\rho)/\partial B_z$ with $D$ when $\alpha = \pi/3$, $B_z = 1$ and $r = 0.5$. (c) The change of $\partial C^2(\rho)/\partial B_z$ with $r$ when $\alpha = \pi/3$, $B_z = 1$ and $D = 0.5$.

Then we can get the change of $\dfrac{\partial C^2(\rho)}{\partial B_z}$ with $B_z$ for $\alpha = \pi/3$, $D = 0.5$ and $r = 0.5$ shown in Fig.4(a), with $D$ for $\alpha = \pi/3$, $B_z = 1$ and $r = 0.5$ in Fig.4(b) and with $r$ for $\alpha = \pi/3$, $B_z = 1$ and $D = 0.5$ in Fig.4(c). From Fig.4, we know $C(\rho)$ is monotone decreasing with magnetic field $B_z$.

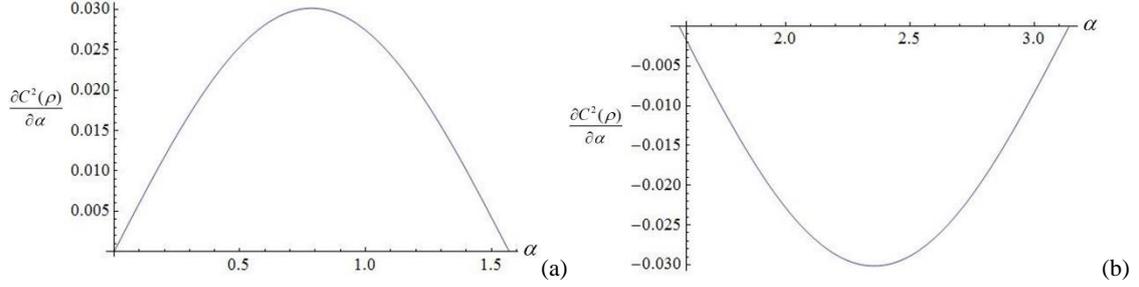

Fig. 5: $\alpha$ dependence of $\partial C^2(\rho)/\partial \alpha$ for $\alpha \in [0, \dfrac{\pi}{2}]$ plotted in (a) and for $\alpha \in [\dfrac{\pi}{2}, \pi]$ plotted in (b) with $B_z = 1$, $D = 0.5$ and $r = 0.5$.

Moreover, we plot $\dfrac{\partial C^2(\rho)}{\partial \alpha}$ of $\alpha \in [0, \dfrac{\pi}{2}]$ and $\alpha \in [\dfrac{\pi}{2}, \pi]$ shown in Fig.5 respectively, for $B_z = 1$, $D = 0.5$ and $r = 0.5$. From Fig.5(a) and (b), we can get $\dfrac{\partial C^2(\rho)}{\partial \alpha} \geq 0$ for $\alpha \in [0, \dfrac{\pi}{2}]$ and $\dfrac{\partial C^2(\rho)}{\partial \alpha} \leq 0$ for $\alpha \in [\dfrac{\pi}{2}, \pi]$, so we know $C(\rho)$ is monotone increasing for $\alpha \in [0, \dfrac{\pi}{2}]$ and monotone decreasing for $\alpha \in [\dfrac{\pi}{2}, \pi]$.

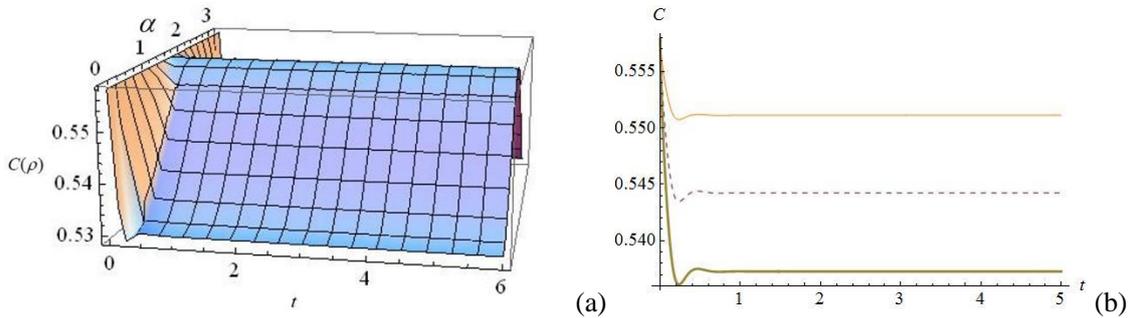

Fig. 6: The change of coherence $C(\rho)$ with $\alpha$ and $t$ plotted in (a) and with $\alpha = \pi/3$ in orange solid line, $\alpha = \pi/4$ in dashed purple line and $\alpha = \pi/6$ in grey thick line plotted in (b).

In what follows, the time evolution of the coherence is investigated; The dynamics of coherence as functions of time $t$ and the initial state parameter $\alpha$ are plotted in Fig.6 (a). Coherence as function of time $t$ for $\alpha = \pi/3$ in orange solid line, $\alpha = \pi/4$ in dashed purple line and $\alpha = \pi/6$ in grey thick line is plotted in Fig.6(b). From Fig.6 (a) and (b), the coherence decreases sharply to a stable value for large value of $\alpha$ and decays slightly for small value of $\alpha$, where $\alpha \in [0, \frac{\pi}{2}]$, and the result will be converse for $\alpha \in [\frac{\pi}{2}, \pi]$. The dynamics of coherence as functions of time $t$ for the intrinsic decoherence parameter $\gamma = 0.05$ in orange solid line, $\gamma = 0.1$ dashed purple line and $\gamma = 0.15$ in thick grey line is plotted in Fig.7. We can find that as the decoherence parameter $\gamma$ increases the coherence decreases more quickly, and there is a oscillation at the bottom of coherence about the balance state.

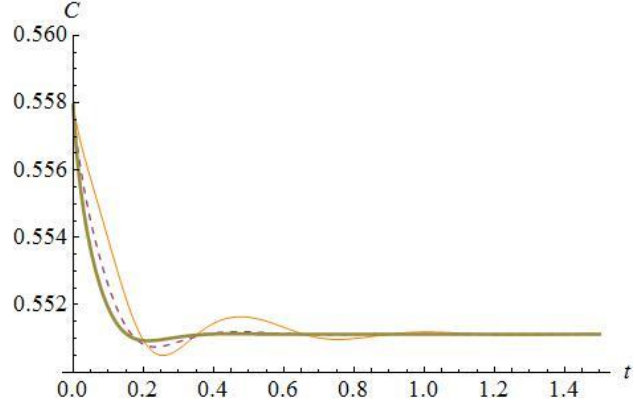

Fig. 7: The change of coherence for time $t$ with the intrinsic decoherence parameter $\gamma = 0.05$ in orange solid line, $\gamma = 0.1$ in dashed purple line and $\gamma = 0.15$ in thick grey line.

## IV. Conclusion

In this paper, we discuss the dynamics of coherence in a two-qubit Heisenberg XXX model with magnetic field in the presence of the intrinsic decoherence. In the long time limit, coherence is monotone increasing with dipole parameter $D$, monotone decreasing with spin distance $r$ and magnetic field $B_z$, and monotone increasing with initial state parameter $\alpha \in [0, \frac{\pi}{2}]$ and monotone decreasing with $\alpha \in [\frac{\pi}{2}, \pi]$. The coherence decreases sharply to a stable value for large value of the initial state parameter $\alpha$ and decays slightly for small value of $\alpha$, where $\alpha \in [0, \frac{\pi}{2}]$, and the result will be converse for $\alpha \in [\frac{\pi}{2}, \pi]$. Moreover, we find that larger decoherence parameter $\gamma$ more quickly the coherence decreases, and there is an oscillation at the bottom of coherence about the balance state.

Our contribution of the study for the current literature is the discussion of coherence of the system with magnetic field, especially with dipole parameter and spin distance. At the same time, it is found that the dipole parameter helps maintain the coherence of a two-qubit Heisenberg system, which is benefit for the future research on decoherence.

## V. Acknowledgements

This work was supported by NSFC under grants Nos.12074027.